\documentclass[mathleft
]{an}
\usepackage{graphicx}
\usepackage{times}
\begin{document}

\Pagespan{789}{}
\Yearpublication{2006}%
\Yearsubmission{2005}%
\Month{11}%
\Volume{999}%
\Issue{88}%

\title{Solar type III bursts with high-frequency cut-off}

\author{A.A. Stanislavsky\thanks{\email{alexstan@ri.kharkov.ua}\newline}
}
\titlerunning{Solar type III bursts with high-frequency cut-off}
\authorrunning{A.A. Stanislavsky}
\institute{Institute of Radio Astronomy, 4, Mystetstv St.,
61002, Kharkiv, Ukraine}

\received{data} \accepted{data} \publonline{later}


\abstract{%
New results in the study of solar type III bursts observed with the UTR-2 radio telescope are presented. The main feature of these bursts is a high-frequency cut-off. The solar activity manifestation was connected with the emergency of a new group of solar spots behind the solar limb relative to an observer on the Earth. This burst type was identified by analyzing its frequency drift rate, duration and flux depending on frequency. The solar bursts were linked to a group of similar events. The cut-off frequency is different from burst to burst and lies within 30-55 MHz. The cut-off origin is considered in the context of propagation effects between the burst sources moving behind the solar limb and the ground-based radio instruments.
}

\maketitle

\section{Introduction}

Solar bursts play an important role in manifestation of solar activity. The radio emission demonstrates a wide variety of properties in time, frequency bandwidth, flux, angular size and polarization which are caused by the complexity and diversity of astrophysical processes responsible for solar radio events. Since the fifties of the last century, their research continues until now together with the intensive development of new modern radio telescopes such as LOFAR, GURT, LWA and others (see, e.\,g., Konovalenko 2005, Ellingson 2005, Stanislavsky et al. 2014 and references therein). The type III bursts are the most population in solar bursts and have been studied most thoroughly both from observational and theoretical points of view (Kr{\"u}ger 1979). Such bursts are observed in a very wide frequency range, from  several GHz down to a few tens of kHz. Their origin is assumed to be fast electron beams propagating with velocities of about 0.3$c$ ($c$ denotes the velocity of light) along open magnetic field lines in the solar corona. The fast electrons induce Langmuir waves along the beam propagation path, and the waves are scattered on ions and transformed into radio emission (see more details, e.\,g., in Zheleznyakov 1970). Nevertheless, even this kind of bursts gives surprises in the study of their observational characteristics. Particularly, some parameters of type III bursts with moderate fluxes differ from similar ones of powerful (flux 1000 s.f.u. and more, recall 1 s.f.u = 10$^{-22}$ W m$^{-2}$ Hz$^{-1}$) type III bursts (see Melnik et al. 2011). Here in this paper, we consider the solar bursts with high-frequency cut-off. Although the solar bursts of this shape were first noted in the 70s (Alvarez et al. 1972), their nature is still not clear until now. Therefore, the analysis of their properties is of great interest for understanding the appearance of such features.

The Earth's atmosphere possesses a low-frequency cut-off (at about 10 MHz) due to the ionosphere which reflects radio waves with frequencies below the plasma frequency of an ionospheric layer. If the type III bursts are observed on spacecrafts (such as ESA-GEOS or Wind/WAVES), their low-frequency cut-off gives an indication of magnetosheath plasma densities. This is important for the case when the plasma density contains regions of abnormally high-density cold plasma (Jones \& Grard 1976). The high-frequency cut-off of solar bursts has some other reasons for the appearance, being connected with coronal processes and observed infrequently. Namely, such a radio event was described in Melnik et al. (2014). It was recorded at frequencies within 16-27.5 MHz on 3 June 2011 at 12:10 UT simultaneously with the UTR-2 and URAN-2 radio telescopes, during the storm of type III bursts. The burst had an unusual fine structure looking like a ``caterpillar''. Although this event was not like any ordinary solar type III burst (in particular, its frequency drift rate was 500 kHz s$^{-1}$ at frequencies higher than 22 MHz and -100 kHz s$^{-1}$ at lower frequencies), this does not mean that it cannot be attributed to the type III bursts. Our observations on 19 August 2012 showed that among solar bursts with high-frequency cut-off there were the events having clear features typical for the well-known type of solar bursts. The point is that the manifestation of solar activity (flares, bursts and others) occurs over the whole Sun, but most of radio astronomy observations are made from the Earth's surface, whereas a significant part of solar radio events (those from the far side of the Sun) is not available for terrestrial observers. To solve this problem, an important contribution is made by two Solar TErrestrial RElations Observatory (STEREO) spacecrafts going in opposite directions along earth orbit around the Sun starting from early 2007. Recently, Krupar et al. (2014) have statistically studied 152 type III radio bursts obtained by the STEREO/Waves observations between May 2007 and February 2013. In about 60\% of events, there were both high- and low-frequency cut-offs because of propagation effects. It will be observed also that the space-based observatories (STEREO and Wind) have too simple antennas for radio observations, and they cannot detect the same number of events as the modern ground-based radio telescopes do (Stanislavsky et al. 2009). Unfortunately, the simultaneous solar radio studies of the whole Sun and its corona is impossible so far. Nevertheless, the solar bursts from the far side of the Sun can arrive at ground-based telescopes due to some specific effects of radio wave propagation. Here, new results of observations are presented at 9-33 MHz for the group of solar type III bursts with high-frequency cut-off, which were observed with the UTR-2 decameter wavelength radio telescope together with other radio astronomy instruments.

\section{Solar phenomena accompanied by solar bursts with high-frequency cut-off}

The solar activity on 17-19 August 2012 was characterized by the emergency of a new group of solar spots near NOAA AR 11548 (N19E78 at 22:46 UT on 18 August 2012) on the far side of the Sun with respect to the ground-based observations. This was due to the generation of many moderate flares over the eastern limb. 19 August 2012 was the final day for the series of repeated flares, when NOAA AR 11548 was on the near side of the Sun. The repeated X-ray flares from the Sun in the first half of that day are shown in Figure~1. A similar form of X-ray solar emission was noticed on 17-18 August.  Moreover, they were accompanied with repeated coronal mass ejections (CMEs).

\begin{figure}
\centering
\resizebox{1.\columnwidth}{!}{%
  \includegraphics{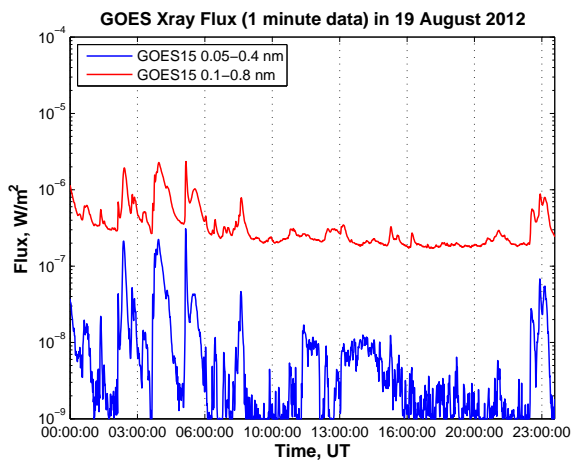}
} \caption{(Color online) A time series of solar X-ray emission observed with the GOES-15 on 19 August 2012.} \label{fig1}
\end{figure}

\section{Facilities}
During that observation, the UTR-2 radio telescope (Braude et al. 1978) was observing in the antenna mode consisting of four sections of the north-south array. This instrument is located near Kharkiv, Ukraine. The total effective area of this antenna part is 50 000 m$^2$ with the beam pattern size of $1^\circ\times 15^\circ$ at 25 MHz. The solar radio data were recorded by the digital DSP spectrometer operating within 9-33 MHz with the time resolution of 100 ms and the frequency resolution of 4 kHz. This device has recently been designed and constructed at the Institute of Radio Astronomy of the National Academy of Sciences of Ukraine (Ryabov et al. 2010). Unfortunately, in our study, the cut-off frequency in the group of bursts (see Figure~2) was close enough to the upper boundary frequency of observations at the UTR-2. The cut-off frequency in some of these bursts was out of the frequency range of this instrument. Therefore, to support these observations, we have used the solar radio data obtained by other radio telescopes suitable for the analysis of these events.

One of them is located at San Vito Solar Observatory (Italy). It is a part of the Radio Solar Telescope Network (RSTN) that consists of ground-based observatories in Australia, Italy, Massachusetts, New Mexico and Hawaii. The solar observatories have a low band (25 to 75 MHz) antenna (non-tracking semi-bicone) and a high band (75 to 180 MHz) antenna (tracking log-periodic) connected to the solar radio spectrograph sweeping the frequency range 25 to 180 MHz every 3 seconds (Coffey 2004). Such an instrument serves to monitor solar radio emissions originating mainly in the solar corona. Although the antenna system of RSTN has a very small effective collecting area and poor sensitivity for reception of weak solar signals, the group of type III bursts under consideration was strong enough in intensity to be recorded by the San Vito Solar Observatory.

Additionally the Nan\c{c}ay Decametric Array (NDA) in France was useful for this study, as it has also received these bursts. The radio telescope has 2 $\times$ 72 helical spiral antennas at 10-80 MHz  for polarization measurements (Boischot et al. 1980). In this work, we took into account the solar NDA data freely available at http://bass2000.obspm.fr/home.php. Despite the difference in the specifications of these radio astronomy tools, scattered world-wide, all these radio records are similar and indicate that the high-frequency cut-off in the bursts had a solar origin rather than ionospheric and/or equipment ones. Both dynamic radio spectra (recorded in San Vito and Nan\c{c}ay) of the solar bursts are shown in Figure~3.

\section{Observations and their analysis}

To study the properties of the solar bursts with high-frequency cut-off, we make use of the radio data obtained with the UTR-2 radio telescope. The data have the highest sensitivity and frequency-time resolution in this frequency range. Figure~2 shows the dynamic radio spectrum during the observation on 19 August 2012. The spectral record represents a group of solar bursts morphologically like the type III bursts. To see this in full, we have analyzed their frequency drift rate $df/dt$ with frequency. For this purpose the maxima of these bursts are taken in time and frequency. This results in 14 sets of hundreds of points which we approximated by the formula
\begin{equation}
f(t) = a(t-b)^{-\alpha}\,, \label{eq1}
\end{equation}
where $a$, $b$ and $\alpha$ are parameters of this model. Each pattern belongs to an individual burst (see the bottom panel of Figure~2). The representation has a clear interpretation. If the emission source moves in the solar corona with constant velocity $v_0$ and negligible acceleration, it travels a distance equal to $r(t) =r_0+v_0(t-t_0)$, where $r_0$ is the initial position, $t_0$ being the starting time (see Lobzin et al. 2008 and Cairns et al. 2009). The background electron density at the source location can be characterized by the local plasma frequency $f_p\propto (r/R_s-1)^{-\alpha}$, where $R_s$ is the solar radius. Then, the source produces radio emission with the frequency drift in time, i.\,e. $f(t) = mA[r_0+v_0(t-t_0)]^{-\alpha}$, where $m=1$ and $m=2$ for the fundamental and harmonic radio emissions of the given burst, respectively. Consequently, we obtain Eq.(\ref{eq1}) in which constants $a$ and $b$ are dependent on constants $A$, $r_0$, $t_0$, $v_0$ and $m$. Next, let us write the frequency drift rate as
\begin{equation}
\frac{df}{dt} = - B\,f^\nu\,,\label{eq2}
\end{equation}
where $B$ and $\nu$ are constants. Note that the values are found directly from the recorded spectra of type III bursts. By the differentiation of Eq.(\ref{eq1}) and the change of variables, we derive the relationship between the constants of Eqs.(\ref{eq1})-(\ref{eq2}), viz. $B=\alpha a^{-1/\alpha}$ and $\nu=1+1/\alpha$.
On the other hand, the equation for the drift rate is
\begin{displaymath}
\frac{df}{dt}=\frac{df}{dn_e}\,\frac{dn_e}{dr}\,v_b=f\frac{1}{2n_e}\,\frac{dn_e}{dr}\,v_b\,,
\end{displaymath}
where $v_b=0.3c$ is the electron beam velocity usually supposed for standard type III bursts, $n_e$ describes the background electron density at the location of the beam. From this follows that $f =v_b^\alpha/(R_s B)^\alpha (r/R_s-1)^{-\alpha}$. Using the relations, we determine the parameters $B$ and $\nu$ for each patterns of the solar bursts (see Table~1). Based on their statistical study, Eq.(\ref{eq2}) takes the form $df/dt = - (0.01 \pm  0.008)\,f^{1.83 \pm 0.39}$. Recall that Alvarez and Haddock (1973) obtained the similar equation $df/dt = - 0.01\,f^{1.84}$ for many solar type III bursts within 75 kHz to 550 MHz. A more recent study of drift rates was made by Mann et al. (1999) who have found $df/dt=-0.0074\,f^{1.76}$. Note that, according to our flux measurements, the solar bursts analyzed were moderate in intensity, as they only approached 1000 s.f.u. The cut-off frequency in the group of bursts varied from burst to burst in the given group and lies within 30-55 MHz, but nothing can be said about their polarization because the UTR-2 radio telescope design is not suitable for polarization measurements. It should be added that the group of solar bursts was associated with the flare of class C2.2, starting at 8:03 UT, having a peak at 8:23 UT and ending at 8:40 UT.

\begin{figure}
\centering
\resizebox{1.\columnwidth}{!}{%
  \includegraphics{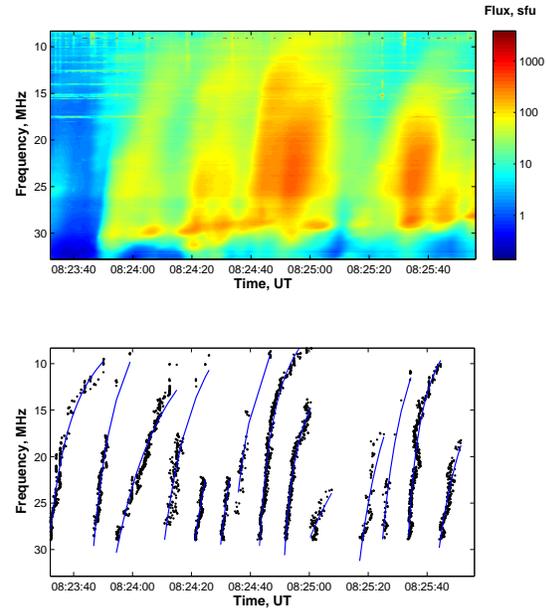}
}  \caption{(Color online) Dynamic spectrum (top panel) of the solar bursts observed on 19 August 2012 at decameter wavelengths with the UTR-2 radiospectrograph and identified as type III bursts. The bottom panel indicates time-frequency traces of peaks in these bursts.} \label{fig2}
\end{figure}

\begin{figure}
\centering
\resizebox{1.\columnwidth}{!}{%
    \includegraphics[width=0.8\textwidth]{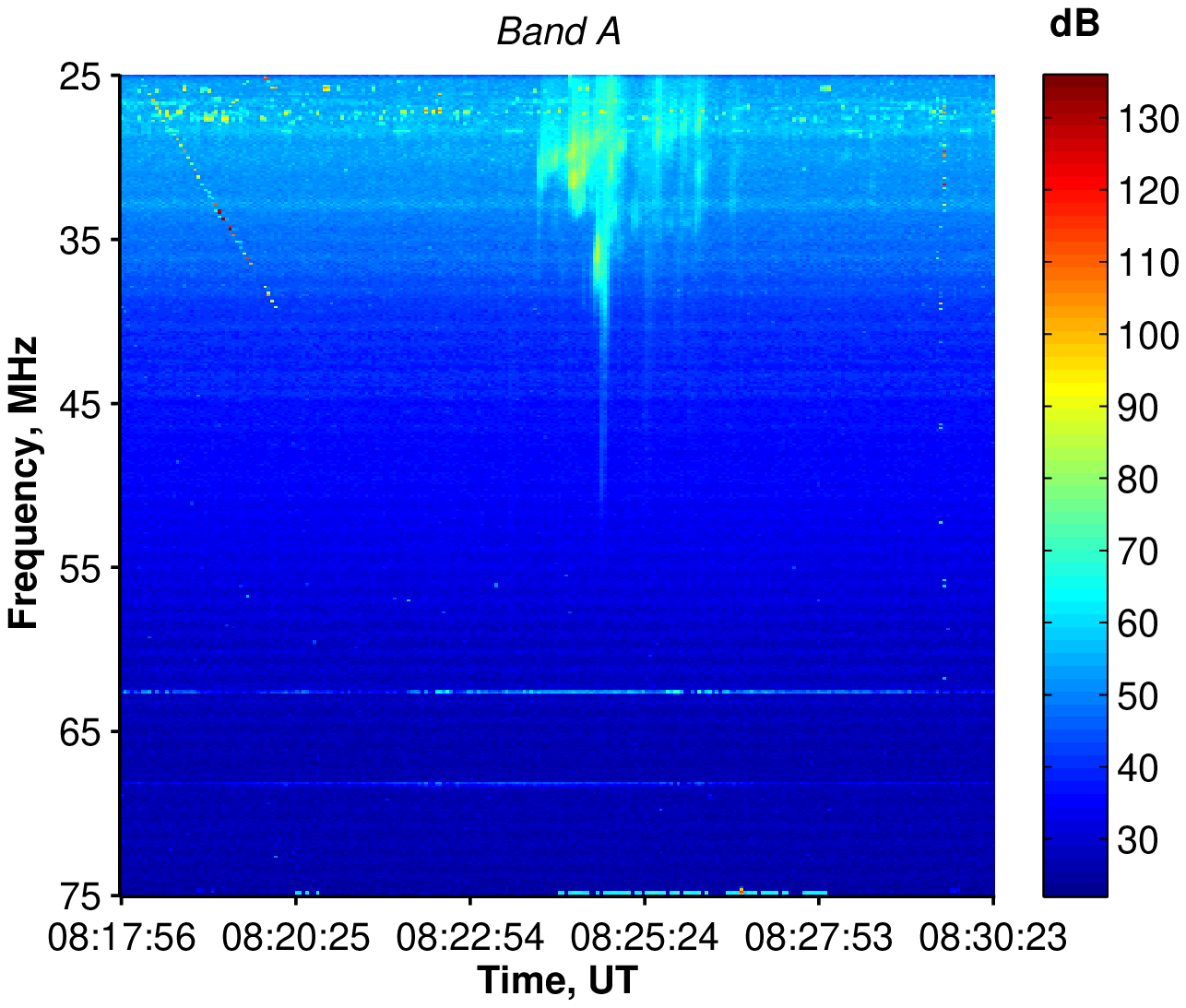}
}   \resizebox{1.\columnwidth}{!}{%
    \includegraphics[width=0.8\textwidth]{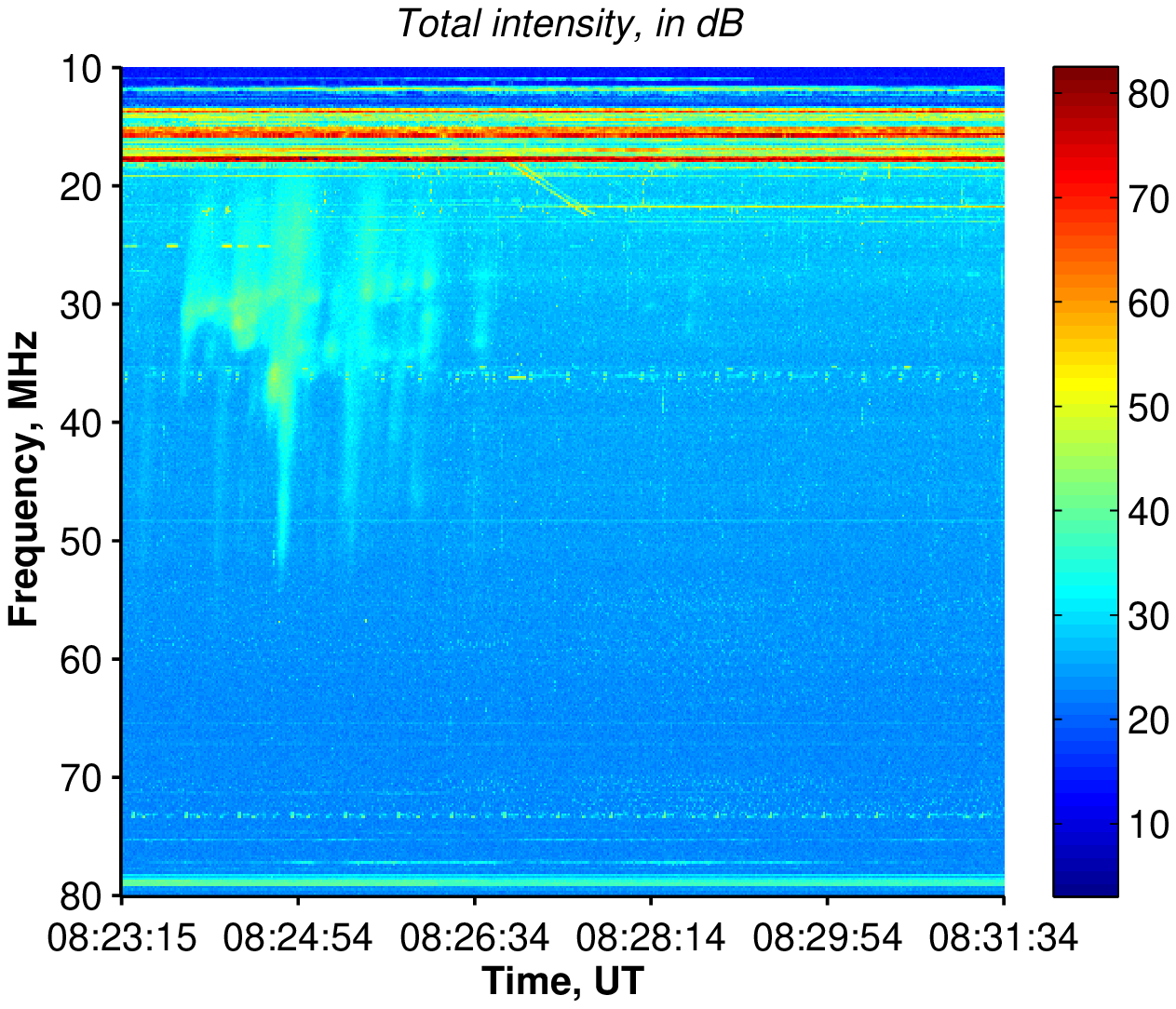}
}
\caption{Solar bursts with high-frequency cut-off observed on 19 August 2012. The records are related to the UTR-2 data shown in Figure~\ref{fig2}. The top panel shows the RSTN data, and the bottom panel does the NDA data.}
    \label{fig3}
\end{figure}

\begin{center}
\begin{table}
\caption[]{Frequency-time parameters of Eq.(\ref{eq2}) as applied to the solar bursts shown in Figure~2.\\
} \label{table2}
\centering
\begin{tabular}{|c|cc|cc|}
\hline
Patterns & $B$ & $\nu$ & $b$ & $\mu$\\
\hline 1 & 0.0033 &	2.0511 & 128.1 & 0.6332 \\
2 & 0.0211 &	1.5115 & 139.6 & 0.6602 \\
3 & 0.0102	& 1.4790 & 195.2 & 0.7698 \\
4 & 0.0145	& 1.5331 & 157.2 & 0.6987 \\
5 & 0.0060 &	1.7658 & 41.3 & 0.2795 \\
6 & 0.0073 & 1.7638 & 34.4 & 0.2219 \\
7 & 0.0233 & 1.5001 & 139.7 & 0.6640 \\
8 & 0.0088 & 1.8783 & 167.4 & 0.7174 \\
9 & 0.0003 & 2.8284 & 160.8 & 0.7049 \\
10 & 0.0003 & 2.3344 & 25.9 & 0.1355 \\
11 & 0.0108 & 1.5843 & 155.5 & 0.6910 \\
12 & 0.0249 & 1.4683 & 57.1 & 0.3794 \\
13 & 0.0052 & 2.0719 & 150.5 & 0.6838 \\
14 & 0.0050 & 1.7980 & 131.6 & 0.6444 \\
\hline
\end{tabular}
\end{table}
\end{center}

Usually, the duration of type III bursts increases as frequency decreases (Wild 1950). Elgar{\o}y and Lyngstad (1972) have analyzed the duration of type III bursts in a wide frequency range, from 300 kHz to 500 MHz. Their best fit between observing frequency $f$ and duration $\tau$ leads to the following dependence
\begin{equation}
\tau = b\,f^{-\mu}\label{eq3}
\end{equation}
with $b=60$, $\mu = 2/3$. Unfortunately, in our case the type III bursts had a strong tendency to occur in a group. Therefore, it is not easy to find the duration of each burst in the whole band of observations because of their overlapping. Table~2 indicates the cases where we were able to determine the half-flux duration $\tau$ depending on $f$. Using the data set, the averaged relationship $\tau(f)$ is written as $\tau = (120.3 \pm 55.8)/f^{0.56 \pm 0.21}$ that is consistent with the results of Elgar{\o}y and Lyngstad (1972). Thus, our analysis of frequency drift rates and durations in solar bursts shows that the group of solar bursts at $\sim$ 08:24 UT on 19 August 2012 has clear signs typical for well-known type III bursts (Reid and Ratcliffe 2014).

\section{Discussion}
Now we discuss how the high-frequency cut-off in solar bursts could be produced. For the cut-off emergence there are several ways: changing a directivity of radiation, intrinsic features of radiation mechanism and propagation effects between the source and the observer. As in the upper corona, where solar radio emission at the decameter wavelength range originates, the local plasma frequency is significantly greater than the local electron gyrofrequency, the magnetic field strength at these heights in solar corona is not enough to strongly influence the burst sources (electron beams) and deflect their radiation away from the Earth. On the other hand, it is hardly possible to imagine that the actual source region of accelerated electrons suddenly was born so high in the corona. The point is that acceleration of solar electrons is attributed to reconfiguration of unstable coronal magnetic fields, and this phenomenon generally occurs at solar active regions (Kr{\"u}ger 1979) rather than in the upper corona. As for radio wave propagation effects between the source and the observer, they can serve as one of preferable mechanisms to explain the high-frequency cut-off in solar bursts.

In this case, the simplest interpretation of solar bursts with high-frequency cut-off is based on assumption that their radiating sources move behind the solar limb relative to an observer on Earth. If the bursts are emitted at the second harmonic of the local plasma frequency, then the high-frequency part of solar burst radiation has not passed by the solar corona in the direction towards the Earth. This idea was used for interpreting the high-frequency cut-off in the event occurred on 3 June 2011. The study was supported by the polarization measurements with the help of the URAN-2 radio telescope (see Melnik et al. 2014). Interestingly, this event had about 10\% level of polarization typical for the second harmonic emission in the type III bursts. However, it should be remembered about the importance of refraction of radio waves in the solar corona for the events of this kind.

\begin{figure}
\centering
\resizebox{1.\columnwidth}{!}{%
  \includegraphics{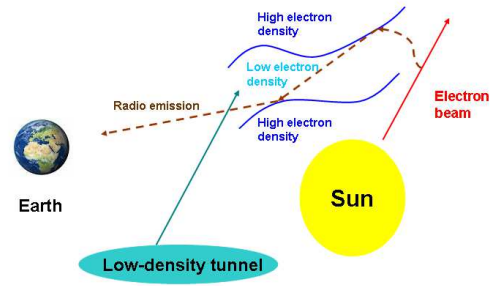}
}  \caption{(Color online) Sketch of ray tracing in solar corona with a low-density cavity due to a CME. Solar limb-behind bursts are generated by electron beams.} \label{fig4}
\end{figure}

The path of a radio emission ray in the solar corona obeys the Snell's law and depends on variations of the refractive index with height. Using the approach of Bracewell and Preston (1956) it is easy to show that if the plasma density monotonically decreases in the solar corona with distance from photosphere, then the radiation of limb-behind burst sources cannot be received by the ground-based radio astronomy instruments in any way due to the radio emission reflection in the direction out of the earth. But if the solar corona contains tunnel-like cavities with low density, the radiation is refracted in corona rather than reflected. In this case, the radiation can be received by terrestrial observers (Figure~4). As is shown by Gibson et al. (2010), such a form of cavities arises before the occurrence of CMEs and erupts into space as a CME abruptly. The higher the actual brightness in CME pieces, the higher the density there. Now recall that the radiation is produced by Thompson's scattering of photospheric light by free electrons, and its brightness is proportional to electron density. If a CME emerges, it has typically a lower-density cavity behind its front (dark bands at SOHO LASCO frames). Moreover, such a cavity often contains a bright core. Under favorable conditions, this permits solar burst sources, moving behind the cavity, to send their radio signals towards the Earth due to the refractive index variations. This concept has been verified successfully by computer simulations, and it will be presented elsewhere later. The cut-off edge is not sharp in Figure 3 that can be explained by changing the cavity density during the CME development.

The acceleration of solar electrons, leading to generation of type III bursts, is possible as well, when the magnetic field in a coronal hole interacts with the surrounding magnetic field (Reid and Ratcliffe 2014). Reverting to the observations of 19 August 2012, it will be observed that the coronal hole (ID 293340 from the Heliophysics Feature Catalogue) observed by the space observatory SDO/AIA/AIA was just on the solar limb behind NOAA AR 11548. Moreover, on the far side of the Sun, several active regions were detected from the STEREO Behind observations. Namely, they appeared on the solar disk as the groups of spots NOAA AR 11555, 11551 and 11554 on 25 August 2012. Probably, the flare activity occurred on the far side of the Sun, and radiation of solar behind-limb bursts could be redirected to the Earth due to the propagation effects between the burst sources and the ground-based observer because of the impact of CMEs near the solar limb. The structure and dynamics of the solar atmosphere is shaped not only by the large eruptions evolving into CMEs, but smaller eruptions, too.  They can cover also a burst source at low altitudes and result in the high-frequency cut-off of solar bursts. According to Eruption Patrol (Hurlburt 2015), the suitable eruption from 8:00 UT to 8:40 UT was observed on 19 August 2012 near NOAA AR 11548 (see, e.\,g., at http://www.helioviewer.org/, as well as at http://spaceweather.gmu.edu/seeds/dailymkmovie.php? cme=20120819).

\section{Conclusions}
Based on this study, we have shown that the solar bursts (observed on 19 August 2012) with high-frequency cut-off are nothing else but the type III radio bursts. This is not a surprise because such bursts are the most numerous events in solar activity. Secondly, we have found the real confirmation in the radio data. This group of solar bursts has frequency drift rates and durations of individual bursts similar to the features of type III radio bursts at low frequencies. The solar bursts, observed on 17-18 August 2012, had also high-frequency cut-offs (Brazhenko et al. 2015), but they were looking less like the type III radio bursts in time-frequency properties. Nevertheless, their properties can be explained in terms of propagation effects between the burst sources and the observer. The authors connected the cut-off appearance in the solar bursts with the solar activity on the far side of the Sun and near the solar limb by virtue of NOAA AR 11548. Note that because of inhomogeneities in tunnel-like cavities of the solar corona, the effects of radio wave propagation (diffraction, multipath, scattering and others) can distort the observed properties of the bursts with high-frequency cut-off. Something similar was apparently observed in the event on 3 June 2011 (Melnik et al. 2014), which had an unusual fine structure with filaments, and its frequency drift rate was noticeably atypical (too slow) for ordinary type III bursts, but notably higher than frequency drift rates of the type II bursts. \\

The author thanks the GOES, NDA and RSTN Teams for designing, constructing and operating the instruments, as well as for their open data policy. Also, I am very grateful to the anonymous Referee for the evaluation of my paper and for the constructive critics. I would like to express my gratitude to A. Kyrychenko for his offering many suggestions to make the final manuscript more readable. This research effort was partially supported by Research Grant 0115U004085 from the National Academy of Sciences of Ukraine.

\end{document}